\input harvmac
\input amssym
\input epsf

\def\S{{\bf S}}



\newfam\frakfam
\font\teneufm=eufm10
\font\seveneufm=eufm7
\font\fiveeufm=eufm5
\textfont\frakfam=\teneufm
\scriptfont\frakfam=\seveneufm
\scriptscriptfont\frakfam=\fiveeufm


\def\bb{
\font\tenmsb=msbm10
\font\sevenmsb=msbm7
\font\fivemsb=msbm5
\textfont1=\tenmsb
\scriptfont1=\sevenmsb
\scriptscriptfont1=\fivemsb
}



\newfam\dsromfam
\font\tendsrom=dsrom10
\textfont\dsromfam=\tendsrom
\def\ds{\fam\dsromfam \tendsrom}


\newfam\mbffam
\font\tenmbf=cmmib10
\font\sevenmbf=cmmib7
\font\fivembf=cmmib5
\textfont\mbffam=\tenmbf
\scriptfont\mbffam=\sevenmbf
\scriptscriptfont\mbffam=\fivembf


\newfam\mbfcalfam
\font\tenmbfcal=cmbsy10
\font\sevenmbfcal=cmbsy7
\font\fivembfcal=cmbsy5
\textfont\mbfcalfam=\tenmbfcal
\scriptfont\mbfcalfam=\sevenmbfcal
\scriptscriptfont\mbfcalfam=\fivembfcal


\newfam\mscrfam
\font\tenmscr=rsfs10
\font\sevenmscr=rsfs7
\font\fivemscr=rsfs5
\textfont\mscrfam=\tenmscr
\scriptfont\mscrfam=\sevenmscr
\scriptscriptfont\mscrfam=\fivemscr




\def\tilde{\widetilde}

\def\bar{\overline}
\def\b{\bar}
\def\bsq#1{{{\b{#1}}^{\lower 2.5pt\hbox{$\scriptstyle 2$}}}}
\def\bexp#1#2{{{\b{#1}}^{\lower 2.5pt\hbox{$\scriptstyle #2$}}}}
\def\dotexp#1#2{{{#1}^{\lower 2.5pt\hbox{$\scriptstyle #2$}}}}


\def\rt2{\sqrt{2}}

\def\grad{\nabla}

\def\Tr{\mathop{\rm Tr}}


\font\tenbifull=cmmib10
\font\tenbimed=cmmib7
\font\tenbismall=cmmib5
\textfont9=\tenbifull \scriptfont9=\tenbimed
\scriptscriptfont9=\tenbismall

\mathchardef\bbGamma="7000
\mathchardef\bbDelta="7001
\mathchardef\bbPhi="7002
\mathchardef\bbAlpha="7003
\mathchardef\bbXi="7004
\mathchardef\bbPi="7005
\mathchardef\bbSigma="7006
\mathchardef\bbUpsilon="7007
\mathchardef\bbTheta="7008
\mathchardef\bbPsi="7009
\mathchardef\bbOmega="700A
\mathchardef\bbalpha="710B
\mathchardef\bbbeta="710C
\mathchardef\bbgamma="710D
\mathchardef\bbdelta="710E
\mathchardef\bbepsilon="710F
\mathchardef\bbzeta="7110
\mathchardef\bbeta="7111
\mathchardef\bbtheta="7112
\mathchardef\bbiota="7113
\mathchardef\bbkappa="7114
\mathchardef\bblambda="7115
\mathchardef\bbmu="7116
\mathchardef\bbnu="7117
\mathchardef\bbxi="7118
\mathchardef\bbpi="7119
\mathchardef\bbrho="711A
\mathchardef\bbsigma="711B
\mathchardef\bbtau="711C
\mathchardef\bbupsilon="711D
\mathchardef\bbphi="711E
\mathchardef\bbchi="711F
\mathchardef\bbpsi="7120
\mathchardef\bbomega="7121
\mathchardef\bbvarepsilon="7122
\mathchardef\bbvartheta="7123
\mathchardef\bbvarpi="7124
\mathchardef\bbvarrho="7125
\mathchardef\bbvarsigma="7126
\mathchardef\bbvarphi="7127


\def\alphadot{{\dot\alpha}}
\def\betadot{{\dot\beta}}




\def\CA{{\cal A}}

\def\CJ{{\cal J}}
\def\CK{{\cal K}}
\def\CL{{\cal L}}
\def\CM{{\cal M}}
\def\CN{{\cal N}}
\def\CO{{\cal O}}

\def\CR{{\cal R}}


\def\1{{\ds 1}}
\def\R{\hbox{$\bb R$}}

\def\S{\hbox{$\bb S$}}


\def\CM{{\cal M}}

\def\K3{{\bf K3}}
\def\journal#1&#2(#3){\unskip, \sl #1\ \bf #2 \rm(19#3) }
\def\andjournal#1&#2(#3){\sl #1~\bf #2 \rm (19#3) }

\def\bar{\overline}

\def\tilde{\widetilde}

\def\frac#1#2{{#1\over#2}}

\def\inbar{\,\vrule height1.5ex width.4pt depth0pt}
\def\IC{\relax\hbox{$\inbar\kern-.3em{\rm C}$}}
\def\IR{\relax{\rm I\kern-.18em R}}
\def\IP{\relax{\rm I\kern-.18em P}}

%
%

%
\catcode`\@=11
\def\slash#1{\mathord{\mathpalette\c@ncel{#1}}}
\overfullrule=0pt

\def\underrel#1\over#2{\mathrel{\mathop{\kern\z@#1}\limits_{#2}}}

\catcode`\@=12


%

\def\exp{{\rm exp}}

\def\unit{\relax{\rm 1\kern-.26em I}}
\def\nada{\relax{\rm 0\kern-.30em l}}
\def\tilde{\widetilde}

\def\alphadot{{\dot \alpha}}
\def\betadot{{\dot\beta}}


\noblackbox
\def\IL{\relax{\rm I\kern-.18em L}}
\def\IH{\relax{\rm I\kern-.18em H}}
\def\IR{\relax{\rm I\kern-.18em R}}
\def\IC{\relax\hbox{$\inbar\kern-.3em{\rm C}$}}
\def\IZ{\relax\ifmmode\mathchoice
{\hbox{\cmss Z\kern-.4em Z}}{\hbox{\cmss Z\kern-.4em Z}} {\lower.9pt\hbox{\cmsss Z\kern-.4em Z}}
{\lower1.2pt\hbox{\cmsss Z\kern-.4em Z}}\else{\cmss Z\kern-.4em Z}\fi}
\def\CM {{\cal M}}
\def\CN {{\cal N}}
\def\CR {{\cal R}}

\def\CJ {{\cal J}}

\def\CL {{\cal L}}

\def\CO {{\cal O}}

\def\CA{{\cal A}}
\def\CK{{\cal K}}
\def\CM {{\cal M}}
\def\CN {{\cal N}}

\def\CO {{\cal O}}

\def\Tr{{\rm Tr}}

\font\manual=manfnt \def\dbend{\lower3.5pt\hbox{\manual\char127}}

\def\IZ{\relax\ifmmode\mathchoice
{\hbox{\cmss Z\kern-.4em Z}}{\hbox{\cmss Z\kern-.4em Z}} {\lower.9pt\hbox{\cmsss Z\kern-.4em Z}}
{\lower1.2pt\hbox{\cmsss Z\kern-.4em Z}}\else{\cmss Z\kern-.4em Z}\fi}

\def\alphadot{{\dot \alpha}}
\def\betadot{{\dot \beta}}

\def\bar{\overline}

\def\rt2{\sqrt{2}}
\def\irt2{{1\over\sqrt{2}}}

\def\slashchar#1{\setbox0=\hbox{$#1$}           
   \dimen0=\wd0                                 
   \setbox1=\hbox{/} \dimen1=\wd1               
   \ifdim\dimen0>\dimen1                        
      \rlap{\hbox to \dimen0{\hfil/\hfil}}      
      #1                                        
   \else                                        
      \rlap{\hbox to \dimen1{\hfil$#1$\hfil}}   
      /                                         
   \fi}

\def\foursqr#1#2{{\vcenter{\vbox{
    \hrule height.#2pt
    \hbox{\vrule width.#2pt height#1pt \kern#1pt
    \vrule width.#2pt}
    \hrule height.#2pt
    \hrule height.#2pt
    \hbox{\vrule width.#2pt height#1pt \kern#1pt
    \vrule width.#2pt}
    \hrule height.#2pt
        \hrule height.#2pt
    \hbox{\vrule width.#2pt height#1pt \kern#1pt
    \vrule width.#2pt}
    \hrule height.#2pt
        \hrule height.#2pt
    \hbox{\vrule width.#2pt height#1pt \kern#1pt
    \vrule width.#2pt}
    \hrule height.#2pt}}}}
\def\psqr#1#2{{\vcenter{\vbox{\hrule height.#2pt
    \hbox{\vrule width.#2pt height#1pt \kern#1pt
    \vrule width.#2pt}
    \hrule height.#2pt \hrule height.#2pt
    \hbox{\vrule width.#2pt height#1pt \kern#1pt
    \vrule width.#2pt}
    \hrule height.#2pt}}}}
\def\sqr#1#2{{\vcenter{\vbox{\hrule height.#2pt
    \hbox{\vrule width.#2pt height#1pt \kern#1pt
    \vrule width.#2pt}
    \hrule height.#2pt}}}}

\def\figin{\epsfcheck\figin}\def\figins{\epsfcheck\figins}
\def\epsfcheck{\ifx\epsfbox\UnDeFiNeD
\message{(NO epsf.tex, FIGURES WILL BE IGNORED)}
\gdef\figin##1{\vskip2in}\gdef\figins##1{\hskip.5in}
\else\message{(FIGURES WILL BE INCLUDED)}%
\gdef\figin##1{##1}\gdef\figins##1{##1}\fi}
\def\DefWarn#1{}
\def\figinsert{\goodbreak\midinsert}
\def\ifig#1#2#3{\DefWarn#1\xdef#1{fig.~\the\figno}
\writedef{#1\leftbracket fig.\noexpand~\the\figno}%
\figinsert\figin{\centerline{#3}}\medskip\centerline{\vbox{\baselineskip12pt \advance\hsize by
-1truein\noindent\footnotefont{\bf Fig.~\the\figno:\ } \it#2}}
\bigskip\endinsert\global\advance\figno by1}



\lref\CremmerEN{
  E.~Cremmer, S.~Ferrara, L.~Girardello, A.~Van Proeyen,
  ``Yang-Mills Theories with Local Supersymmetry: Lagrangian, Transformation Laws and SuperHiggs Effect,''
Nucl.\ Phys.\  {\bf B212}, 413 (1983).
}

\lref\DUSE{T.~Dumitrescu and N.~Seiberg, to appear}

\lref\SUSYADS{O.~Aharony and N.~Seiberg, unpublished. }

\lref\WessCP{
  J.~Wess, J.~Bagger,
  ``Supersymmetry and supergravity,''
Princeton, USA: Univ. Pr. (1992) 259 p.
}

\lref\StelleYE{
  K.~S.~Stelle, P.~C.~West,
  ``Minimal Auxiliary Fields for Supergravity,''
Phys.\ Lett.\  {\bf B74}, 330 (1978).
}

\lref\FerraraEM{
  S.~Ferrara, P.~van Nieuwenhuizen,
  ``The Auxiliary Fields of Supergravity,''
Phys.\ Lett.\  {\bf B74}, 333 (1978).
}

\lref\SohniusTP{
  M.~F.~Sohnius, P.~C.~West,
  ``An Alternative Minimal Off-Shell Version of N=1 Supergravity,''
Phys.\ Lett.\  {\bf B105}, 353 (1981).
}

\lref\SohniusFW{
  M.~Sohnius, P.~C.~West,
  ``The Tensor Calculus And Matter Coupling Of The Alternative Minimal Auxiliary Field Formulation Of N=1 Supergravity,''
Nucl.\ Phys.\  {\bf B198}, 493 (1982).
}

\lref\GiudiceXP{
  G.~F.~Giudice, M.~A.~Luty, H.~Murayama, R.~Rattazzi,
  ``Gaugino mass without singlets,''
JHEP {\bf 9812}, 027 (1998).
[hep-ph/9810442].
}

\lref\AdamsVW{
  A.~Adams, H.~Jockers, V.~Kumar and J.~M.~Lapan,
  ``N=1 Sigma Models in $AdS_4$,''
  arXiv:1104.3155 [hep-th].
}

\lref\RandallUK{
  L.~Randall, R.~Sundrum,
  ``Out of this world supersymmetry breaking,''
Nucl.\ Phys.\  {\bf B557}, 79-118 (1999).
[hep-th/9810155].
}

\lref\DineME{
  M.~Dine, N.~Seiberg,
  ``Comments on quantum effects in supergravity theories,''
JHEP {\bf 0703}, 040 (2007).
[hep-th/0701023].
}

\lref\KomargodskiPC{
  Z.~Komargodski, N.~Seiberg,
  ``Comments on the Fayet-Iliopoulos Term in Field Theory and Supergravity,''
JHEP {\bf 0906}, 007 (2009).
[arXiv:0904.1159 [hep-th]].
}

\lref\GripaiosRG{
  B.~Gripaios, H.~D.~Kim, R.~Rattazzi, M.~Redi, C.~Scrucca,
  ``Gaugino mass in AdS space,''
JHEP {\bf 0902}, 043 (2009).
[arXiv:0811.4504 [hep-th]].
}

\lref\ImamuraUW{
  Y.~Imamura,
  ``Relation between the 4d superconformal index and the $S^3$ partition function,''
[arXiv:1104.4482 [hep-th]].
}

\lref\FerraraPZ{
  S.~Ferrara, B.~Zumino,
  ``Transformation Properties of the Supercurrent,''
Nucl.\ Phys.\  {\bf B87}, 207 (1975).
}

\lref\GatesNR{
  S.~J.~Gates, M.~T.~Grisaru, M.~Rocek, W.~Siegel,
  ``Superspace Or One Thousand and One Lessons in Supersymmetry,''
Front.\ Phys.\  {\bf 58}, 1-548 (1983).
[hep-th/0108200].
}

\lref\AharonyAY{
  O.~Aharony, D.~Marolf, M.~Rangamani,
  ``Conformal field theories in anti-de Sitter space,''
JHEP {\bf 1102}, 041 (2011).
[arXiv:1011.6144 [hep-th]].
}

\lref\GirardiVQ{
  G.~Girardi, R.~Grimm, M.~Muller and J.~Wess,
  ``Antisymmetric Tensor Gauge Potential In Curved Superspace And A (16+16)
  Supergravity Multiplet,''
  Phys.\ Lett.\  B {\bf 147}, 81 (1984).
}

\lref\LangXK{
  W.~Lang, J.~Louis and B.~A.~Ovrut,
  ``(16+16) Supergravity Coupled To Matter: The Low-Energy Limit Of The
  Superstring,''
  Phys.\ Lett.\  B {\bf 158}, 40 (1985).
}

\lref\SiegelSV{
  W.~Siegel,
  ``16/16 Supergravity,''
  Class.\ Quant.\ Grav.\  {\bf 3}, L47 (1986).
}

\lref\KomargodskiRB{
  Z.~Komargodski, N.~Seiberg,
  ``Comments on Supercurrent Multiplets, Supersymmetric Field Theories and Supergravity,''
JHEP {\bf 1007}, 017 (2010).
[arXiv:1002.2228 [hep-th]].
}

\lref\PestunRZ{
  V.~Pestun,
  ``Localization of gauge theory on a four-sphere and supersymmetric Wilson loops,''
[arXiv:0712.2824 [hep-th]].
}

\lref\PestunNN{
  V.~Pestun,
  ``Localization of the four-dimensional N=4 SYM to a two-sphere and 1/8 BPS Wilson loops,''
[arXiv:0906.0638 [hep-th]].
}

\lref\OkudaKE{
  T.~Okuda, V.~Pestun,
  ``On the instantons and the hypermultiplet mass of N=2* super Yang-Mills on $S^{4}$,''
[arXiv:1004.1222 [hep-th]].
}

\lref\RomelsbergerEG{
  C.~Romelsberger,
  ``Counting chiral primaries in N = 1, d=4 superconformal field theories,''
Nucl.\ Phys.\  {\bf B747}, 329-353 (2006).
[hep-th/0510060].
}

\lref\RomelsbergerEC{
  C.~Romelsberger,
  ``Calculating the Superconformal Index and Seiberg Duality,''
[arXiv:0707.3702 [hep-th]].
}

\lref\DolanQI{
  F.~A.~Dolan, H.~Osborn,
  ``Applications of the Superconformal Index for Protected Operators and q-Hypergeometric Identities to N=1 Dual Theories,''
Nucl.\ Phys.\  {\bf B818}, 137-178 (2009).
[arXiv:0801.4947 [hep-th]].
}

\lref\KinneyEJ{
  J.~Kinney, J.~M.~Maldacena, S.~Minwalla, S.~Raju,
  ``An Index for 4 dimensional super conformal theories,''
Commun.\ Math.\ Phys.\  {\bf 275}, 209-254 (2007).
[hep-th/0510251].
}

\lref\SenPH{
  D.~Sen,
  ``Supersymmetry In The Space-time R X S**3,''
Nucl.\ Phys.\  {\bf B284}, 201 (1987).
}

\lref\JafferisUN{
  D.~L.~Jafferis,
  ``The Exact Superconformal R-Symmetry Extremizes Z,''
[arXiv:1012.3210 [hep-th]].
}

\lref\HamaAV{
  N.~Hama, K.~Hosomichi, S.~Lee,
  ``Notes on SUSY Gauge Theories on Three-Sphere,''
JHEP {\bf 1103}, 127 (2011).
[arXiv:1012.3512 [hep-th]].
}

\lref\SpiridonovZA{
  V.~P.~Spiridonov, G.~S.~Vartanov,
  ``Elliptic hypergeometry of supersymmetric dualities,''
[arXiv:0910.5944 [hep-th]].
}

\lref\KapustinXQ{
  A.~Kapustin, B.~Willett, I.~Yaakov,
  ``Nonperturbative Tests of Three-Dimensional Dualities,''
JHEP {\bf 1010}, 013 (2010).
[arXiv:1003.5694 [hep-th]].
}

\lref\KapustinMH{
  A.~Kapustin, B.~Willett, I.~Yaakov,
  ``Tests of Seiberg-like Duality in Three Dimensions,''
[arXiv:1012.4021 [hep-th]].
}

\lref\KapustinGH{
  A.~Kapustin,
  ``Seiberg-like duality in three dimensions for orthogonal gauge groups,''
[arXiv:1104.0466 [hep-th]].
}

\lref\JafferisZI{
  D.~L.~Jafferis, I.~R.~Klebanov, S.~S.~Pufu, B.~R.~Safdi,
  ``Towards the F-Theorem: N=2 Field Theories on the Three-Sphere,''
[arXiv:1103.1181 [hep-th]].
}

\lref\WillettGP{
  B.~Willett, I.~Yaakov,
  ``N=2 Dualities and Z Extremization in Three Dimensions,''
[arXiv:1104.0487 [hep-th]].
}

\lref\SeibergVC{
  N.~Seiberg,
  ``Naturalness versus supersymmetric nonrenormalization theorems,''
Phys.\ Lett.\  {\bf B318}, 469-475 (1993).
[hep-ph/9309335].
}

\lref\HamaEA{
  N.~Hama, K.~Hosomichi, S.~Lee,
  ``SUSY Gauge Theories on Squashed Three-Spheres,''
[arXiv:1102.4716 [hep-th]].
}

\lref\CheonVI{
  S.~Cheon, H.~Kim, N.~Kim,
  ``Calculating the partition function of N=2 Gauge theories on $S^3$ and AdS/CFT correspondence,''
[arXiv:1102.5565 [hep-th]].
}

\lref\MartelliQJ{
  D.~Martelli, J.~Sparks,
  ``The large N limit of quiver matrix models and Sasaki-Einstein manifolds,''
[arXiv:1102.5289 [hep-th]].
}

\lref\AmaritiHW{
  A.~Amariti,
  ``On the exact R charge for N=2 CS theories,''
[arXiv:1103.1618 [hep-th]].
}

\lref\DrukkerZY{
  N.~Drukker, M.~Marino, P.~Putrov,
  ``Nonperturbative aspects of ABJM theory,''
[arXiv:1103.4844 [hep-th]].
}

\lref\MarinoNM{
  M.~Marino,
  ``Lectures on localization and matrix models in supersymmetric Chern-Simons-matter theories,''
[arXiv:1104.0783 [hep-th]].
}

\lref\DolanRP{
  F.~A.~H.~Dolan, V.~P.~Spiridonov, G.~S.~Vartanov,
  ``From 4d superconformal indices to 3d partition functions,''
[arXiv:1104.1787 [hep-th]].
}

\lref\CheonTH{
  S.~Cheon, D.~Gang, S.~Kim, J.~Park,
  ``Refined test of AdS4/CFT3 correspondence for N=2,3 theories,''
[arXiv:1102.4273 [hep-th]].
}

\lref\HerzogHF{
  C.~P.~Herzog, I.~R.~Klebanov, S.~S.~Pufu, T.~Tesileanu,
  ``Multi-Matrix Models and Tri-Sasaki Einstein Spaces,''
Phys.\ Rev.\  {\bf D83}, 046001 (2011).
[arXiv:1011.5487 [hep-th]].
}

\lref\NiarchosSN{
  V.~Niarchos,
  ``Comments on F-maximization and R-symmetry in 3D SCFTs,''
[arXiv:1103.5909 [hep-th]].
}

\lref\MinwallaMA{
  S.~Minwalla, P.~Narayan, T.~Sharma, V.~Umesh, X.~Yin,
  ``Supersymmetric States in Large N Chern-Simons-Matter Theories,''
[arXiv:1104.0680 [hep-th]].
}

\lref\SenBG{
  D.~Sen,
  ``Extended Supersymmetry In The Space-time R X S**3,''
Phys.\ Rev.\  {\bf D41}, 667 (1990).
}

\lref\KeckSE{
  B.~W.~Keck,
  ``An Alternative Class of Supersymmetries,''
J.\ Phys.\ A {\bf A8}, 1819-1827 (1975).
}

\lref\ZuminoAV{
  B.~Zumino,
  ``Nonlinear Realization of Supersymmetry in de Sitter Space,''
Nucl.\ Phys.\  {\bf B127}, 189 (1977).
}

\lref\SpiridonovZR{
  V.~P.~Spiridonov, G.~S.~Vartanov,
  ``Superconformal indices for N = 1 theories with multiple duals,''
Nucl.\ Phys.\  {\bf B824}, 192-216 (2010).
[arXiv:0811.1909 [hep-th]].
}

\lref\GaddeTE{
  A.~Gadde, L.~Rastelli, S.~S.~Razamat, W.~Yan,
  ``The Superconformal Index of the $E_6$ SCFT,''
JHEP {\bf 1008}, 107 (2010).
[arXiv:1003.4244 [hep-th]].
}

\lref\SpiridonovHH{
  V.~P.~Spiridonov, G.~S.~Vartanov,
  ``Supersymmetric dualities beyond the conformal window,''
Phys.\ Rev.\ Lett.\  {\bf 105}, 061603 (2010).
[arXiv:1003.6109 [hep-th]].
}

\lref\VartanovXJ{
  G.~S.~Vartanov,
  ``On the ISS model of dynamical SUSY breaking,''
Phys.\ Lett.\  {\bf B696}, 288-290 (2011).
[arXiv:1009.2153 [hep-th]].
}

\lref\ButterYM{
  D.~Butter and S.~M.~Kuzenko,
  ``N=2 AdS supergravity and supercurrents,''
  arXiv:1104.2153 [hep-th].
}

\lref\IvanovVB{
  E.~A.~Ivanov, A.~S.~Sorin,
  ``Superfield Formulation Of Osp(1,4) Supersymmetry,''
J.\ Phys.\ A {\bf A13}, 1159-1188 (1980).
}

\lref\IvanovFT{
  E.~A.~Ivanov, A.~S.~Sorin,
  ``Wess-Zumino Model as Linear Sigma Model of Spontaneously Broken Conformal and OSp(1,4) Supersymmetries,''
Sov.\ J.\ Nucl.\ Phys.\  {\bf 30}, 440 (1979).
}

\lref\BandosNN{
  I.~A.~Bandos, E.~Ivanov, J.~Lukierski, D.~Sorokin,
  ``On the superconformal flatness of AdS superspaces,''
JHEP {\bf 0206}, 040 (2002).
[hep-th/0205104].
}

\lref\GaddeIA{
  A.~Gadde, W.~Yan,
  ``Reducing the 4d Index to the $\S^3$ Partition Function,''
[arXiv:1104.2592 [hep-th]].
}

\lref\GaddeKB{
  A.~Gadde, E.~Pomoni, L.~Rastelli and S.~S.~Razamat,
  ``S-duality and 2d Topological QFT,''
  JHEP {\bf 1003}, 032 (2010)
  [arXiv:0910.2225 [hep-th]].
}

\lref\GaddeEN{
  A.~Gadde, L.~Rastelli, S.~S.~Razamat and W.~Yan,
  ``On the Superconformal Index of N=1 IR Fixed Points: A Holographic Check,''
  JHEP {\bf 1103}, 041 (2011)
  [arXiv:1011.5278 [hep-th]].
}

\lref\FRS{G.~Festuccia, M.~Rocek and N.~Seiberg, to appear.}


\Title{
} {\vbox{\centerline{Rigid Supersymmetric Theories}
\centerline{}
\centerline{ in Curved Superspace}
}}
\medskip

\centerline{\it Guido Festuccia and Nathan Seiberg}
\bigskip
\centerline{School of Natural Sciences}
\centerline{Institute for Advanced Study}
\centerline{Einstein Drive, Princeton, NJ 08540}

\smallskip

\vglue .3cm

\bigskip
\noindent
We present a uniform treatment of rigid supersymmetric field theories in a curved spacetime $\CM$, focusing on four-dimensional theories with four supercharges.  Our discussion is significantly simpler than earlier treatments, because we use classical background values of the auxiliary fields in the supergravity multiplet.  We demonstrate our procedure using several examples.  For $\CM=AdS_4$ we reproduce the known results in the literature.  A supersymmetric Lagrangian for $\CM=\S^4$ exists, but unless the field theory is conformal, it is not reflection positive.  We derive the Lagrangian for $\CM=\S^3\times \R$ and note that the time direction $\R$ can be rotated to Euclidean signature and be compactified to $\S^1$ only when the theory has a continuous R-symmetry.  The partition function on $\CM=\S^3\times \S^1$ is independent of the parameters of the flat space theory and depends holomorphically on some complex background gauge fields.  We also consider R-invariant $\CN=2$ theories on $\S^3$ and clarify a few points about them.

 \Date{05/2011}


\newsec{Introduction}

Recently, different lines of investigations have focused on supersymmetric field theories on spheres.  Pestun~\refs{\PestunRZ,\PestunNN} computed the partition function and the expectation value of circular Wilson loops in $\CN=4$ and some $\CN=2$ theories on $\S^4$ (see also~\refs{\OkudaKE}). In three dimensions Kapustin, Willet and Yaakov~\refs{\KapustinXQ\KapustinMH\KapustinGH-\WillettGP } used localization techniques to compute the partition function of several $\CN=2$ theories on $\S^3$ as a mean to test certain conjectured dualities. This work has been followed by~\refs{\HamaEA\CheonVI\HamaAV\MartelliQJ\DrukkerZY\HerzogHF-\MarinoNM} and inspired Jafferis to propose ``$Z$-minimization'' \refs{\JafferisUN}, which has spurred several studies \refs{\JafferisZI\AmaritiHW\CheonTH\NiarchosSN-\MinwallaMA}. Some of these three-dimensional $\CN=2$ theories are related by dimensional reduction to $\CN=1$ theories on $\S^3\times \R$.  Such theories were originally studied by Sen \refs{\SenPH,\SenBG} and more recently by Romelsberger \refs{\RomelsbergerEG,\RomelsbergerEC}, who used $\CN=1$ theories on $\S^3\times \S^1$  to define an index which reduces to~\refs{\KinneyEJ} when the theory is superconformal. This index has been computed for different theories~\refs{\DolanQI\SpiridonovZR\SpiridonovZA\GaddeKB\GaddeTE\GaddeEN
\SpiridonovHH\VartanovXJ-\DolanRP}  providing checks of several dualities.

Our starting point is the flat space Lagrangian $\CL_{\R^4}$ written in terms of the component dynamical fields, which include the auxiliary fields in the matter and gauge multiplets.  We would like to replace flat $\R^4$ with a curved space $\CM$.  Depending on certain properties of $\CM$, which we will discuss below, the theory on $\CM$ can be supersymmetric.  In general, the supersymmetry generators on $\CM$ are a subset of those on $\R^4$ and furthermore their algebra is deformed.

We would like to understand the conditions on $\CM$ such that it admits a supersymmetry algebra, to identify this algebra and to find the deformation $\CL_{\CM}$ of $\CL_{\R^4}$ such that this supersymmetry algebra is preserved.  A first attempt to find $\CL_{\CM}$ is simply to introduce the metric into $\CL_{\R^4}$.  In general the resulting Lagrangian $\CL_{\CM}^0$ is not supersymmetric.  We can correct it by adding a power series in $1\over r$
\eqn\curvLag{\CL_{\CM}=\CL^{(0)}_{\CM} +\delta \CL_{\CM}=\sum_{n=0}^\infty {1\over r^n}\CL^{(n)}_{\CM}~,}
where $r$ is the characteristic size of $\CM$, defined by scaling the metric $g_{\mu \nu}= r^2 g_{\mu \nu}^{(0)}$.  Note that there are two sources of $r$ dependence in \curvLag.  First, there is $r$ dependence in the metric which appears in all the terms $\CL_{\CM}^{(n)}$.  Second, we have the explicit factors of ${1\over r}$ in the coefficients.  In the rest of this note we will determine the correction terms $\CL_{\CM}^{(n)}$.  Surprisingly, we will find that they vanish for $n>2$.

One approach to finding the algebra and the Lagrangian is to start with $\CL^{(0)}_{\CM}$ and to derive the corrections to the supersymmetry algebra, the supersymmetry variation of the fields and the Lagrangian by a perturbation expansion in $1\over r$.  This approach is clearly correct, but it is technically complicated and it is not clear that the expansion in $1\over r$ would terminate.

Instead, we will describe an alternate procedure, which makes the construction of the theory straightforward.  In the spirit of \SeibergVC\ we will couple the theory to classical background fields and will promote them to superfields.  For a theory in curved spacetime we need to specify the metric as well as the  {\it auxiliary fields} in the gravity multiplet. It is important that unlike the ordinary use of auxiliary fields, we do not solve their equations of motion -- we specify arbitrary values for them.  In some of the examples below we will also add background gauge fields.  Furthermore, we will find it necessary to let some of these background fields be complex even though they must be real in ``sensible theories.''

Before we proceed, we would like to make two general comments about placing theories in curved space.  First, given a flat space Lagrangian, the curved space Lagrangian is always ambiguous.  There can be terms that vanish in the flat space limit because they multiply powers of the curvature.  In addition, we can always add terms to the flat space theory that are multiplied by additional parameters like the overall scale of the metric.  For example, when we put a flat space theory on a sphere with radius $r$ we have freedom in adding arbitrary terms of order $1\over r$.  Below we will encounter such ambiguities.  Our main concern here will be to determine the terms that {\it must be added to the flat space theory in order to preserve supersymmetry}.

Second, standard low energy effective Lagrangian techniques, which are extremely powerful when the theory is in flat space, might not be applicable here.  Normally, we integrate out high momentum modes and expand the effective Lagrangian in low momenta.  In curved spacetime we cannot integrate out momenta of order the inverse radius of curvature and restrict the effective Lagrangian to the terms with at most two derivatives.  The reason for that is that there is no invariant way to separate higher derivative terms in the effective Lagrangian from terms that are suppressed by the inverse radius of curvature.

The outline of the rest of the paper is as follows. In section 2 we explain the general procedure; starting from off shell supergravity we take a rigid limit holding the metric and auxiliary fields of the gravity multiplet fixed to values constrained only by the requirement that the resulting theory has some degree of supersymmetry. Our approach differs from previous ones in that we do not integrate out the auxiliary fields by using their equations of motion. We comment on some properties of the resulting theories and we identify $\CL^{(1)}_{\CM}$ in \curvLag\ in terms of the Ferrara-Zumino supercurrent multiplet.

In section 3 we apply the general formalism to the well known case of $AdS_4$. We comment on the impossibility to put a theory without an FZ-multiplet in $AdS$ and on the fact that arguments of holomorphy which are very powerful in flat space are not useful in this case. In section 4 we apply our procedure to the case of $\S^4$. The resulting Lagrangian, while supersymmetric, is not reflection positive unless the theory is superconformal.

In section 5 we consider $\S^3\times\R$. We find that in order for the supercharges to be time independent the theory must have an R-symmetry. We also comment on the fact that holomorphy arguments are applicable in this case. Motivated by the R-symmetry requirement, in section 6 we obtain the theories on $\S^3\times \R$ by taking the decoupling limit in ``new minimal'' supergravity \refs{\SohniusTP}. We are then able to identify $\CL^{(1)}_{\CM}$ in the terms of the R-multiplet. This helps in clarifying the structure of the Lagrangians presented in section~5.

In section 7 we consider $\S^3\times\S^1$. We analytically continue the $\S^3\times \R$ construction to Euclidean space and compactify it on a circle. The partition function $Z$ depends on the dimensionless ratio $\beta \over r$ of the size $\beta$ of the circle and the radius $r$ of the sphere.  It also depends on complex background gauge fields along the circle $v_s$, where $s$ labels the global symmetries of the theory.  The dependence of $Z$ on $v_s$ is holomorphic.  Finally, this partition function is independent of all the parameters of the flat space theory.

Section 8 is devoted to R-invariant $\CN=2$ theories on $\S^3$.  Generic such theories are not reflection positive.  Our perspective, which is based on background fields clarifies a number of features of these theories.

\newsec{General Procedure}

Our procedure starts by coupling the flat space theory $\CL_{\R^4}$ to off-shell supergravity.  We assume for simplicity that the theory can be coupled to the ``old minimal set of auxiliary fields''~\refs{\StelleYE,\FerraraEM}.  Below we will comment on the situation in which such a coupling is impossible. In this formalism the graviton multiplet consists of the graviton, the gravitino and some auxiliary fields.  The auxiliary fields are a complex scalar $M$ and a real vector $b_\mu$.  Since the auxiliary fields do not propagate, it is common to integrate them out using their classical equations of motion.  Instead, we prefer to keep them in the Lagrangian.

Next, we want to decouple the fluctuations in the gravitational field such that it remains a classical background.  This is achieved by taking the Planck scale $M_P$ to infinity.  As we do that we have to decide how to scale the various fields.  We assign dimension zero to the metric $g_{\mu \nu}$ and dimension one to $M$ and $b_\mu$.

It is important to stress that this limit is not the same as the linearized supergravity limit.  In the latter we expand around flat space $g_{\mu \nu}= \eta_{\mu \nu}+ {1\over M_P} h_{\mu \nu}$ when we take the Planck scale to infinity.  Instead, we want to keep a nontrivial metric in our limit.

For simplicity we focus on a flat space theory $\CL_{\R^4}$ that is based on chiral superfields with K\"ahler potential $K$ and superpotential $W$.  It is straightforward to add gauge fields to this theory.  Taking the limit we mentioned above in the supergravity Lagrangian~\refs{\CremmerEN,\WessCP}, setting the gravitino to zero and dropping terms that are independent of the dynamical matter superfields we find\foot{We use the conventions of~\refs{\WessCP}, except that we define $v_{\alpha \alphadot}=-2 \sigma^{\mu}_{\alpha \alphadot} v_{\mu}$, so that $v_{\mu}={1\over 4} \bar \sigma_{\mu}^{ \alphadot \alpha }v_{\alpha \alphadot}$.}
\eqn\partsugra{\eqalign{
&\CL=\CL^B+\CL^F \cr
&{1\over e}\CL^B=\left({1\over 6}\CR+{1\over 9} M\bar M-{1\over 9} b_{\mu}b^{\mu}\right)K+ K_{i\bar j}\left( F^i \bar F^{\bar j} -\partial_{\mu} \bar \phi^{\bar j}\partial^{\mu} \phi^i\right)\cr
&\qquad\quad + F^i W_i +\bar F^{\bar j} \bar W_{\bar j}-{1\over 3} K_{i}M F^i -{1\over 3} K_{\bar i}\bar M \bar F^{\bar i} -W \bar M-\bar W M\cr &\qquad\quad-{i\over 3}b^{\mu} \left(K_i \partial_{\mu} \phi^i-K_{\bar i} \partial_{\mu} \bar \phi^{\bar i} \right) \cr
&{1\over e}\CL^F=-{i} K_{i \bar j}\bar \psi^{\bar j} \bar \sigma^{\mu} \tilde \grad_{\mu}\psi^i-{1\over 2} W_{i j} \psi^i \psi ^j -{1\over 2} \bar W_{\bar i \bar j} \bar \psi^{\bar i} \bar \psi^{\bar j}\cr
&\qquad\quad -{1\over 2}K_{i j \bar j} \bar F^{\bar j} \psi^i \psi^j -{1\over 2}K_{\bar i \bar j  j}  F^{ j} \bar \psi^{\bar i} \bar \psi^{\bar j}+{1\over 4}K_{ i j \bar i \bar j} \psi^i \psi^j \bar \psi^{\bar i}\bar \psi^{\bar j}\cr
&\qquad\quad-{1\over 6 } b^{\mu} K_{i\bar i} \psi^i \sigma_{\mu} \bar \psi^{\bar i}+{1\over 6} MK_{i j} \psi^i \psi^j +{1\over 6} \bar M K_{\bar i \bar j} \bar \psi^{\bar i}\bar \psi^{\bar j} ~,}}
where $\CR$ is the scalar curvature (which is negative for a sphere and positive in $AdS$) and
\eqn\covde{\eqalign{
&\tilde \grad_{\mu}\psi^i=\grad_{\mu} \psi^i+\Gamma^{i}_{j l}\psi^j \partial_{\mu}\phi^l \cr
&\Gamma^{i}_{j k}= K^{i \bar i}K_{j k \bar i}~. }}
For reasons that will be important below we allow $b_\mu$ to be complex and $\bar M$ not to be the complex conjugate of $M$.

The freedom in performing K\"ahler transformations of the underlying supergravity Lagrangian might not be preserved by our classical background fields. It is straightforward to check that provided the background satisfies
\eqn\Kahlercond{\eqalign{
&{3\over 2 }\CR - b_\mu b^\mu - 2 M\bar M=0\cr
&\grad_\mu b^\mu =0~,}}
the Lagrangian \partsugra\ is invariant (up to a total derivative) under the K\"ahler transformation
\eqn\Kahlerrig{\eqalign{
&K \to K + Y(\phi) + \bar Y(\bar\phi) \cr
&W\to W+ {1\over 3 }M Y \cr
&\bar W\to \bar W+ {1\over 3}\bar M \bar Y ~.}}

It is convenient to introduce the Ferrara-Zumino supercurrent multiplet~\refs{\FerraraPZ,\GatesNR}:
\eqn\FZmul{\bar D^\alphadot \CJ_{\alpha\alphadot} = D_\alpha X \qquad ;\qquad \bar D_\alphadot X=0.}
For a WZ-model
\eqn\newcomb{\eqalign{
&\CJ_{\mu}|= J^{FZ}_{\mu}={2 i\over 3}\Big(\partial_{\mu} \phi^i K_i-\partial_{\mu}\bar \phi^{\bar i} K_{\bar i}\Big)+{1\over 3 } K_{i\bar i} \psi^i \sigma_{\mu} \bar \psi^{\bar i}
\cr
&X|=4 W - {1\over 3 }\bar D^2 K|= 4 W + {4\over 3} K_{\bar i} \bar F^{\bar i} - {2\over 3} K_{\bar i \bar j} \bar \psi^{\bar i}\bar \psi^{\bar j}~.}}
Therefore, we can interpret the terms linear in the auxiliary fields  $b_{\mu}$ and $M,\bar M$ in \partsugra\ as
\eqn\auxlin{-{1\over 2} b^{\mu} J_{\mu}^{FZ}-{1\over 4} M \bar X|-{1\over 4} \bar M X| ~.}

Even though we derived equation \auxlin\ using the WZ-model, it is valid for all flat space theories that have an FZ-multiplet\foot{See \refs{\KomargodskiRB,\DUSE} for a detailed discussion and a list of earlier references.}.  These include all abstract theories even when no explicit Lagrangian description is possible.  In every such theory the operators $J_\mu^{FZ}$ and $X$ exist and the leading order deformation of the Lagrangian is given by \auxlin.  Below we will comment about the curved space description of theories without an FZ-multiplet.

The FZ-multiplet \FZmul\ is not unique.  It is subject to improvement transformations $X \to X + \bar D^2 \Omega$ with chiral $\Omega$.  In the WZ-model this can be identified with K\"ahler transformations $K\to K - 3(\Omega + \bar \Omega)$ \KomargodskiPC.   If \Kahlercond\ are satisfied, the freedom in $\Omega$ can be absorbed in shifting the rigid theory superpotential as $W\to W -M \Omega$, which reflects the freedom in performing the transformation \Kahlerrig. Note that such a shift of the superpotential is an example of the ambiguity we discussed in the introduction.

The supersymmetry variation of the matter fields is
\eqn\varsm{\eqalign{
&\delta \phi^i= -\sqrt{2} \zeta \psi^i\cr
&\delta \psi^i_{\alpha}= -\sqrt{2} \zeta_{\alpha} F^i-i \sqrt{2} (\sigma^{\mu}\bar \zeta)_{\alpha} \partial_{\mu} \phi^i \cr
&\delta F^i=-i \sqrt{2} \bar \zeta \bar \sigma^{\mu} \grad_{\mu} \psi^i-{\sqrt{2}\over 3}\bar M \zeta\psi^i+{\sqrt{2}\over 6}b_{\mu} \bar \zeta\bar\sigma^{\mu}\psi^{i},  }}
while for the gravitino
\eqn\vars{\eqalign{
&\delta \Psi_{\mu}^{\alpha}=-2 \grad_{\mu} \zeta^{\alpha} +{i\over 3}\left(M (\epsilon \sigma_{\mu} \bar \zeta)^{\alpha}+2 b_{\mu} \zeta^{\alpha} +2b^{\nu} (\zeta \sigma_{\nu\mu})^{\alpha}\right)\cr
&\delta \bar \Psi_{\mu \dot \alpha}=-2 \grad_{\mu} \bar \zeta_{\dot \alpha} -{i\over 3}\left(\bar M (  \zeta \sigma_{\mu} )_{\dot \alpha}+2 b_{\mu} \bar \zeta_{\dot \alpha} +2b^{\nu} ( \bar \zeta \bar \sigma_{\nu \mu})_{\dot \alpha}\right)~.}}
The two equations \vars\ are complex conjugate of each other.  But given that we allow complex $b_{\mu}$ and $M$ and $\bar M$ as independent complex functions, we should impose both of them.

Unlike standard treatments of supergravity, we have not eliminated the auxiliary fields $M$ and $b_\mu$.  Furthermore, we have not performed the customary Weyl rescaling which sets the Einstein-Hilbert term to its canonical form.  Because of these two facts, the variation of the gravitino \vars\ is independent of the matter fields.

Now we are ready to specify a classical background for our rigid theory.  It is characterized by the values of the metric and the auxiliary fields $M$ and $b_\mu$.  It is important that these values are completely arbitrary.  They do not have to satisfy any equations of motion.

If we want our background to be supersymmetric, it should allow nontrivial solutions of
\eqn\susycon{\eqalign{
&\delta \Psi_{\mu}^{\alpha}=-2 \grad_{\mu} \zeta^{\alpha} +{i\over 3}\left(M (\epsilon \sigma_{\mu} \bar \zeta)^{\alpha}+2 b_{\mu} \zeta^{\alpha} +2b^{\nu} (\zeta \sigma_{\nu\mu})^{\alpha}\right)=0\cr
&\delta \bar \Psi_{\mu \dot \alpha}=-2 \grad_{\mu} \bar \zeta_{\dot \alpha} -{i\over 3}\left(\bar M (  \zeta \sigma_{\mu} )_{\dot \alpha}+2 b_{\mu} \bar \zeta_{\dot \alpha} +2b^{\nu} ( \bar \zeta \bar \sigma_{\nu \mu})_{\dot \alpha}\right)=0 ~.}}

Given a background metric $M$, $\bar M$ and $b_\mu$, which satisfy \susycon\ with nonzero $\zeta, \bar \zeta$, our theory has some unbroken supersymmetry.  It arises as a subalgebra of the local super-diffeomorphism of the underlying supergravity theory.  In general, it is different than the rigid flat space supersymmetry algebra we started with.

A detailed analysis of the conditions \susycon\ with various number of unbroken supersymmetries will be presented elsewhere \FRS.  Here we simply state that demanding four unbroken supersymmetries, results in:
\eqn\finalsol{\eqalign{
&Mb_\mu=\bar M b_\mu=0\cr
&\grad_{\mu}b_{\nu} =0\cr
&\partial_\mu M=\partial_\mu \bar M=0 \cr
&W_{\mu\nu\kappa\lambda}=0\cr
&\CR_{\mu\nu} =-{2\over 9} (b_\mu b_\nu - g_{\mu \nu} b_\rho b^\rho)+{1\over 3} g_{\mu\nu}M \bar M}} where $W_{\mu\nu\kappa\lambda}$ is the Weyl tensor. In particular \Kahlercond\ is satisfied. Because the metric is conformally flat the supersymmetry algebra is a subalgebra of the $SU(2,2|1)$ superconformal algebra.

 There are two classes of solutions of \finalsol:
\item{1} $b_{\mu}=0$ with constant $M,\bar M$ these will be considered in sections 3 and 4.
\item{2} $M=\bar M=0$ with $b_{\mu}$ a covariantly constant vector. The metric is conformally flat and further restricted by \finalsol. The case of $\S^3\times \R$ analyzed in sections 5-7 belongs to this class.

 \noindent Furthermore,  we can immediately identify the terms in the expansion \curvLag.  $\CL_\CM^{(0)}$ arises from using the metric in the flat space Lagrangian.  $\CL_\CM^{(1)}$ arises from the terms in \partsugra\ that are linear in the auxiliary fields -- i.e. it arises from \auxlin.  And $\CL_\CM^{(2)}$ arises from the terms in \partsugra\ that are linear in $\CR$ or quadratic in the auxiliary fields.  Below we will see an example in which it is natural to make another field redefinition which leads to additional contributions to $\CL_\CM^{(1,2)}$.  But in all cases it is clear that the expansion \curvLag\ stops at $n=2$.

\newsec{$AdS_4$}

As our first nontrivial example we place the rigid theory in $AdS_4$ with curvature
\eqn\curvAdS{\CR={12\over r^2} }
i.e.\ $r$ is the curvature radius.

Several authors starting with~\refs{\KeckSE\ZuminoAV\IvanovFT-\IvanovVB} have studied supersymmetric field theories in $AdS_4$ (for a recent discussion and a list of earlier references see~\refs{\AdamsVW}).  One approach starts by identifying the supersymmetry algebra $OSp(1|4)$.  Then one iteratively finds the Lagrangian and the supersymmetry transformation laws.  Alternatively, we can simply add a constant $M_p^2\over r$ to the superpotential and solve the gravitational equations of motion to put the system in $AdS_4$.  Then, one can scale $M_p\to \infty$ in order to decouple the gravitational field.  The approach we take here is clearly equivalent to this one but uses the more general procedure of the previous section.

The conditions  for unbroken supersymmetry \susycon\ are satisfied on $AdS_4$ with the choice
\eqn\AdSM{M=\bar M=-{3\over r} \qquad , \qquad b_\mu=0~;}
the supersymmetry parameter $\zeta_{\alpha}$ satisfies:
\eqn\AdSS{2 \grad_{\mu} \zeta^{\alpha} +{i\over r} (\epsilon \sigma_{\mu} \bar \zeta)^{\alpha}=0}
and its complex conjugate.

The terms in the curved space Lagrangian that originate form the background auxiliary fields are
\eqn\extraM{{1\over e } \delta\CL_{AdS} ={3\over r^2 }K +{1\over r}(  K_i F^i +  K_{\bar i} \bar F^{\bar i}) +{3\over r} {W} +{3\over r} {\bar W}
-{1\over 2r}K_{i j} \psi^i \psi^j -{1\over 2r} K_{\bar i \bar j} \bar \psi^{\bar i}\bar \psi^{\bar j} ~,}
where we have used \curvAdS.  The full Lagrangian is
\eqn\lagaux{\eqalign{
&{1\over e}\CL_{AdS}={1\over e}\CL^B_{AdS}+{1\over e}\CL^F_{AdS} \cr
&{1\over e}\CL^B_{AdS}
=-\CK_{i \bar i}\partial_\mu \phi^i \partial^\mu \bar \phi^{\bar i} +\CK_{i \bar i}F^i \bar F^{\bar i} +{1\over  r} \Big(\CK_i F^i + \CK_{\bar i} \bar F^{\bar i} \Big)+{3\over r^2} \CK  \cr
&{1\over e}\CL^F_{AdS}
=- i \CK_{i \bar i} \bar\psi^{\bar i} \bar \sigma^{\mu} \tilde\grad_{\mu} \psi^i -{1\over 2} \CK_{i \bar i \bar j} F^i \bar \psi^{\bar i}\bar \psi^{\bar j}-{1\over 2} \CK_{i j \bar j} \bar F^{\bar j} \psi^{ i} \psi^{ j}+{1\over 4} \CK_{i j \bar i \bar j} \psi^i\psi^j \bar \psi^{\bar i} \bar \psi^{\bar j}
\cr
&\qquad\qquad -{1\over 2r } \Big(\CK_{i j} \psi^i \psi^j+\CK_{\bar i \bar j}\bar \psi^{\bar i}\bar \psi^{\bar j} \Big)\cr
&\tilde \grad_{\mu}\psi^i=\grad_{\mu} \psi^i+\Gamma^{i}_{j l}\psi^j \partial_{\mu}\phi^l \cr
&\Gamma^{i}_{j k}= \CK^{i \bar i}\CK_{j k \bar i}\cr
&\CK=K +  r (W+\bar W)  .}}

Since $AdS$ is conformally flat (see also\BandosNN ), its $OSp(1|4)$ superalgebra is a subalgebra of the flat space superconformal algebra $SU(2,2|1)$.  Its bosonic $Sp(4) \cong SO(3,2)$ subalgebra is the isometry of $AdS_4$ and can be viewed as a deformation of the flat space Poincar\'e symmetry.  The four flat space supersymmetry generators are deformed to be two supersymmetry generators and two superconformal generators from $SU(2,2|1)$.

Since the conditions \Kahlercond\ are satisfied, the Lagrangian is invariant under the K\"ahler transformations \Kahlerrig\
    \eqn\kahlert{\eqalign{
    &K\to K +Y(\phi) +\bar Y (\bar \phi)\cr
    &W \to W - {1\over r} Y (\phi) \cr
    &\bar W \to \bar W - {1\over r} \bar Y (\bar\phi) ~.}}
This explains the dependence of the Lagrangian on $\CK$.

The operator $X$ in \FZmul\ includes the trace of the energy momentum tensor and it reflects the breaking of superconformal invariance.  If the theory is superconformal, $X=0$ (or more precisely, $X = \bar D^2 \Omega$ with chiral $\Omega$) and hence $\CL_{AdS}^{(1)}$ in \curvLag\ vanishes.  Indeed, since $AdS_4$ is conformally flat, it is easy to put any conformal field theory on it \AharonyAY.  In this case the procedure based on supergravity is not needed because the curved space Lagrangian is easily determined using conformal invariance.

An interesting application of this observation is in quantum field theories in which $X$ receives radiative corrections.  For examples, in gauge theories, the anomaly shifts $X$ by a term proportional to $\Tr W_\alpha W^\alpha$ and therefore, $\CL_{AdS}^{(1)}$ receives a one loop correction proportional to a gaugino bilinear.  Such a term was studied in the context of anomaly mediation~\refs{\GiudiceXP,\RandallUK} in~\refs{\DineME,\GripaiosRG}.

Even if we started with an R-invariant rigid theory, the nonzero value of $M$ \AdSM\ violates that R-symmetry, so the theory in $AdS_4$ is not R-invariant.  There are two interesting exceptions to this comment.  First, if the flat space rigid theory we start with is superconformal, then the resulting theory in $AdS_4$ is R-invariant.  As we commented above, in this case $X$ vanishes and the R-breaking term $\CL_{AdS}^{(1)}$ is absent.  Second, if the theory has $\CN=2$ supersymmetry, the operator $X|$ belongs to an $SU(2)_R$ triplet.  It breaks it to $U(1)_R \subset SU(2)_R$, and hence the theory in $AdS$ has this $U(1)_R$ symmetry.  This is closely related to the R-symmetry of the theory on $\S^4$ studied by Pestun~\refs{\PestunRZ,\PestunNN}.

Our discussion applies only to rigid supersymmetric theories with an FZ-multiplet.  We claim that rigid theories without an FZ-multiplet cannot be placed in $AdS_4$ (this was also discussed in~\refs{\ButterYM,\AdamsVW}).  To see that, recall that such theories can be coupled to linearized supergravity only when they have a global R-symmetry or additional dynamical fields are added to them \KomargodskiRB.  However, as we commented above, even if we start with an R-invariant theory, the nonzero value of $M$ \AdSM\ violates the symmetry, thus making the theory in $AdS_4$ inconsistent.  For example, theories whose target space does not have an exact K\"ahler form or theories with FI-terms cannot be placed in $AdS$ while preserving supersymmetry.

Since the K\"ahler form of our theory must be exact, we can always use \kahlert\ to set $W=0$.  This explains why \lagaux\ depends only on $\CK$. Note that it is common in the supergravity literature to use such a transformation to set the field dependent part of the superpotential to zero.  In general, one might criticize this practice, because such a transformation could have singularities and might even be inconsistent, if the K\"ahler form of the target space is nontrivial.  However, in our case it is always possible to redefine $W$ into the K\"ahler potential.

Writing the Lagrangian in terms of $\CK$ rather than in terms of $K$ and $W$ shows that the standard separation of the Lagrangian into $K$ and $W$ is not present here.  Therefore, the holomorphy based techniques for controlling the superpotential are not useful.  This point about supersymmetric field theories in $AdS$ has been realized by various people including \refs{\AdamsVW,\SUSYADS}.

Next we integrate out the auxiliary fields $F^i$, $\bar F^{\bar i}$ in \lagaux\ using their classical equations of motion
\eqn\FbarFs{\eqalign{
& F^{ i} =-g^{i\bar i} \Big(\bar W_{\bar i} + {1 \over  r}K_{\bar i}\Big)+{1\over 2}\Gamma^i_{ j l } \psi^{j}  \psi^{ l}=-{1\over r} g^{i\bar i} \CK_{\bar i}+{1\over 2}\Gamma^i_{ j l } \psi^{j}  \psi^{ l}\cr
&\bar F^{\bar i} =-g^{i\bar i} \Big( W_ i + {1\over  r}K_i \Big)+{1\over 2}\Gamma^{\bar i}_{ \bar j \bar l}\bar \psi^{\bar j} \bar \psi^{\bar l}= -{1\over  r} g^{i\bar i} \CK_i+{1\over 2}\Gamma^{\bar i}_{ \bar j \bar l}\bar \psi^{\bar j} \bar \psi^{\bar l}}}
leading to the potential
\eqn\VAdS{\eqalign{
V_{AdS}(\phi) =& g^{i \bar i}W_i\bar W_{\bar i} + {1\over r}\left( g^{i \bar i}K_i \bar W_{\bar i}+ g^{i \bar i} K_{\bar i} \bar W_i - 3 W-3 \bar W\right) +{1\over r^2} \left(g^{i \bar i} K_i K_{\bar i} - 3K\right)\cr
=&{1\over r^2} \left(g^{i \bar i} \CK_i \CK_{\bar i} - 3\CK\right).}}

The conditions for unbroken supersymmetry are $F^i=0$.  These are $n$ complex equations for $n$ complex variables.  It is easy to show that if these equations are satisfied, the potential \VAdS\ is stationary.

The supersymmetric vacua can be analyzed in an expansion in $1/r$.  If the flat space theory does not break supersymmetry, its vacua are at $\phi_0^i$ satisfying $W_i(\phi_0)=0.$  Then, the condition for unbroken supersymmetry $W_i+{1\over r}K_i=0$ are satisfied by $\phi^i=\phi_0^i +{1\over r}\phi_1^i+ \cdots$ with
\eqn\unbroksusd{\eqalign{
&\phi_1^i=-W^{il}(\phi_0) K_l(\phi_0,\bar\phi_0)\cr
&\bar\phi_1^{\bar i}=-\bar W^{\bar i\bar l}(\bar \phi_0) K_{\bar l}(\phi_0,\bar\phi_0)~,}}
where $W^{il}$ is the inverse of the flat space fermion mass matrix $W_{il}$, which we assume to be invertible.

Alternatively, if we want to preserve one of the flat space supersymmetric expectation values
\eqn\flatspace{\eqalign{
&\langle\phi^i\rangle=\phi_0^i \cr
&\langle\bar\phi^{\bar i}\rangle=\bar\phi_0^{\bar i}}}
which satisfy $W_i(\phi_0)=0$, we can shift the superpotential by terms which vanish in the flat space limit ($ r \to \infty$)
\eqn\shiftW{\eqalign{
&\tilde W= W -{1\over r} K_{\bar i}(\phi_0,\bar \phi_0) \phi^i \cr
&\bar{\tilde  W}= \bar W -{1\over r} K_i(\phi_0,\bar \phi_0) \bar \phi^{\bar i} }}
and then the auxiliary fields equations \FbarFs\ become
\eqn\FbarFss{\eqalign{
& F^{ i} =-g^{i\bar i} \Big(\bar {\tilde W}_{\bar i} + {1 \over r}K_{\bar i}\Big) \cr
&\bar F^{\bar i} =-g^{i\bar i} \Big( \tilde W_ i + {1\over  r}K_i \Big)}}
and they vanish at the flat space value \flatspace.  Note that we can do it for each of the supersymmetric solutions of $W_i=0$, but we cannot do it simultaneously for all of them.

\newsec{$\S^4$}

Next we take the theory to be Euclidean and put it on $\S^4$.  Here the supersymmetry condition is satisfied for
\eqn\supcon{\CR=-{12\over r^2} \qquad,\qquad M=\bar M =-{3i \over r} ~.}
Note that $\bar M$ is not the complex conjugate of $M$.

The Lagrangian can be obtained from the Euclidean version of \lagaux\ by $r\rightarrow -i r$
\eqn\lagauxs{\eqalign{
&\CL_{\S^4}=\CL^B_{\S^4}+\CL^F_{\S^4} \cr
&{1\over e}\CL^B_{\S^4}=\CK_{i \bar i}\partial_\mu \phi^i \partial^\mu \bar \phi^{\bar i} -\CK_{i \bar i}F^i \bar F^{\bar i} -{i\over  r} \Big(\CK_i F^i + \CK_{\bar i} \bar F^{\bar i} \Big)+{3\over r^2} \CK  \cr
&{1\over e}\CL^F_{\S^4}= i \CK_{i \bar i} \bar\psi^{\bar i} \bar \sigma^{\mu} (\grad_{\mu} \psi^i + \Gamma^i_{j k} \partial_{\mu}\phi^j \psi^k)-{1\over 4} \CK_{i j \bar i \bar j} \psi^i\psi^j \bar \psi^{\bar i} \bar \psi^{\bar j}\cr &\qquad\qquad+{1\over 2} \CK_{i \bar i \bar j} F^i \bar \psi^{\bar i}\bar \psi^{\bar j}+{1\over 2} \CK_{i j \bar j} \bar F^{\bar j} \psi^{ i} \psi^{ j}+{i\over 2r } (\CK_{i j} \psi^i \psi^j+\CK_{\bar i \bar j}\bar \psi^{\bar i}\bar \psi^{\bar j} )\cr
&\Gamma^{i}_{j k}= \CK^{i \bar i}\CK_{j k \bar i}\cr
&\CK=K  -i r (W+\bar W)  .}}

Note that this bosonic Lagrangian is not real! This originates from $\bar M$ not being the complex conjugate of $M$ in \supcon.  This is in accord with the well known fact that while we can put supersymmetric theories on $AdS$ space, we cannot put them on $dS$ space.  The theory we find on the sphere (which is the Euclidean version of $dS$ space) is not reflection positive and hence it does not correspond to any unitary field theory in Lorentzian signature space.

An obvious exception to this comment is superconformal field theories in $\S^4$.  Since $\S^4$ is conformally flat, it is clear that the resulting theory is reflection positive.  This fact is visible in \lagauxs.  The terms that violate reflection positivity are proportional to $X| - \bar X|$ and these terms vanish in conformal theories.

Even though we do not discuss it here in detail, it is clear that the same issue with lack of reflection positivity applies to non-conformal $\CN=2$ theories on $\S^4$ \refs{\PestunRZ-\OkudaKE} and on $\S^3$ \refs{\KapustinXQ-\MinwallaMA}.

The unusual reality properties of the theory make the interpretation of the dependence on $\CK$ confusing.  Starting with a flat space theory with a real K\"ahler potential $K$ we cannot use a K\"ahler transformation like  \kahlert\  to remove $W$ -- we could do that only if $\bar Y (\bar \phi)$ in \kahlert\ was not the complex conjugate of $Y(\phi)$.  Therefore, one might hope that the standard separation of the data characterizing the theory into a K\"ahler potential $K$ and a superpotential $W$ could be maintained.  We do not pursue this possibility in this publication.

Next we integrate out the auxiliary fields:
\eqn\FbarFsS{\eqalign{
& F^{ i} =-g^{i\bar i} \Big(\bar W_{\bar i} + {i \over  r}K_{\bar i}\Big)=-{i\over r} g^{i\bar i} \CK_{\bar i}\cr
&\bar F^{\bar i} =-g^{i\bar i} \Big( W_ i + {i\over  r}K_i \Big)= -{i\over  r} g^{i\bar i} \CK_i}}
noting that this solution for $\bar F$ is not the complex conjugate of the solution for $F$.

As in the discussion around \unbroksusd, we can look for supersymmetric solutions
\eqn\Feq{\eqalign{
& F^{ i} =-g^{i\bar i} \Big(\bar W_{\bar i} + {i \over  r}K_{\bar i}\Big)=0\cr
&\bar F^{\bar i} =-g^{i\bar i} \Big( W_ i + {i\over  r}K_i \Big)=0}}
in a power series in $1\over r$.  We expand around flat space supersymmetric solutions satisfying $\bar \phi_0^{\bar i}=\phi_0^{i*}$.  Then,
\eqn\sadloc{\eqalign{
&\phi_{s}^{ i}=\phi_{0}^{ i}+{1\over r}\phi_{1}^{i}+\cdots \cr
&\bar \phi_{s}^{ \bar i}=\bar \phi_{0}^{\bar i}+{1\over r} \bar \phi_{1}^{\bar i}+\cdots }}
Here
\eqn\sadeq{\eqalign{
&W_{i}(\phi_{0})=\bar W_{\bar i}(\bar\phi_{0})=0\cr
&\bar\phi_{0}^{\bar i}=\phi_{0}^{ i*}\cr
&\phi_{1}^{ i}=-iW^{i l}(\phi_0) K_{l}( \phi_0,\bar \phi_0) \cr
&\bar \phi_{1}^{ \bar i}=-i\bar W^{\bar i \bar l}(\bar\phi_0) K_{ \bar l}(\phi_0,\bar\phi_0)~. }}
Note that the supersymmetric solutions $\phi_s$, $\bar\phi_s$ are generically such that $\bar \phi_s$ is not the complex conjugate of $\phi_s$; i.e.\ they are not on the standard integration contour of the flat space theory.

The value of the potential at the saddles is:
\eqn\potsaddle{\eqalign{
V |_s&=-i{3\over r} (W |_s+\bar W |_s)+{3\over r^2} K |_s=\cr
&=-i{3\over r}\left(W(\phi_0)+\bar W(\bar \phi_0)\right)+{3\over r^2}K(\phi_0,\bar\phi_0)+\cdots}}
Note that the higher order corrections to the position of the saddle do not affect the potential at this order.  Its imaginary value is determined by the value of the superpotential at the flat space saddle $\phi_0$ and its real part is determined by the K\"ahler potential at that point.

\newsec{$\S^3\times \R$}

We want to study the theory on $\S^3\times \R$ with the sphere of radius $r$ and
\eqn\strR{\CR=-{6\over r^2} ~.}
The conditions \susycon\ for unbroken supersymmetry can be solved on $\S^3\times\R$
by choosing\foot{The isometry group of $\S^3$ is $SU(2)_l\otimes SU(2)_r$. Changing the sign of $b_0$ corresponds to interchanging the role of the two $SU(2)$ factors in what follows.}
\eqn\bzback{b_0=-{3\over r} \qquad, \qquad M=\bar M=b_i=0 ~.}
The supersymmetry parameter $\zeta_{\alpha}$ then satisfies
\eqn\grazer{\eqalign{
&\partial_t \zeta_{\alpha}+{i \over r}  \zeta_{\alpha}=0\cr
&2 \grad_a \zeta_{\alpha}-{i\over r}   (\sigma_a \bar\sigma_0\zeta)_{\alpha}=0 ~.}}

The effective Lagrangian in this background is obtained by substituting the curved metric and the background auxiliary fields \bzback\ in \partsugra.  The contributions to the Lagrangian due to the background auxiliary fields are:
\eqn\extrab{ {1\over e} \delta\CL_{\S^3\times \R}= -{i\over  r}\left(K_i \partial_{t} \phi^i- K_{\bar i} \partial_{t}\bar \phi^i\right)-{1\over 2r } K_{i\bar i} \psi^i \sigma_{0} \bar \psi^{\bar i} }
Here the $\CO(1/r^2)$ are canceled by using \strR.

In accord with \auxlin\ the terms of order ${1\over r}$ are given by $-{3\over 2}j^{FZ}_0$  where $j^{FZ}_{\mu}$ is the current appearing in the lowest component of the FZ-multiplet \FZmul.  As we remarked above, the expression in terms of the operator $j^{FZ}_{\mu}$ is more general than the particular example of WZ-model we used.  It applies in any field theory including theories without a Lagrangian description.

Since \strR\bzback\ satisfy \Kahlercond, our system is invariant under \Kahlerrig
\eqn\Kahlersp{K\rightarrow K(\phi,\bar \phi)+Y(\phi)+\bar Y(\bar \phi)}
without transforming $W$. The separation of holomorphic data in $W$ from the non-holomorphic $K$ present in flat space continues to hold on $\S^3\times\R$. This is one way to see why, unlike $AdS$, here holomorphy is active and can lead to nontrivial results.

As in $AdS$, since this background is conformally flat, the supersymmetry algebra is a subalgebra of the flat space superconformal algebra $SU(2,2|1)$. It is $SU(2|1)_l\otimes SU(2)_r$.  Its bosonic subalgebra is $SU(2)_l\otimes SU(2)_r \otimes U(1)$ which is the isometry of $\S^3 \times \R$.   Some important commutation relations are\foot{Here, and also below, we could absorb the factors of $r$ in a redefinition of the charges.  We do not do it because this presentation allows us to contract the superalgebra to its flat space version by taking $r$ to infinity.}
\eqn\algspc{\eqalign{&\{Q_{\alpha},{\bar Q}_{\dot\alpha}\}=2 \sigma^{0}_{\alpha \dot\alpha} P_0+{2\over r}\sigma^i_{\alpha \dot \alpha} J^i_l\cr
&\{Q_{\alpha},Q_{\beta}\}=0 \cr
&\{\bar Q_{\alphadot},\bar Q_{\betadot}\}=0 \cr
&[P_0,Q_{\alpha}]={1\over r} Q_{\alpha}}}
where $P_0$ generates translations along $\R$, while the $J^i_l$ are the generators for the $SU(2)_l$ subgroup of the $\S^3$ isometries.

The vanishing of the second and third anti-commutators in \algspc\ underlies the fact that the theory on $\S^3\times \R$ has a standard holomorphic superpotential.  This is related to the invariant separation into a K\"ahler potential and a superpotential we mentioned above and is behind the control we have in analyzing such theories.

It is important that the supercharges in this subalgebra do not commute with the generator $P_0$ of translations along $\R$ and hence they are time dependent.  This can be changed, if the theory has an R-symmetry $[R,Q_{\alpha}]=-Q_{\alpha}$.  Denoting the R-charges of $\phi^i$ by $q_i$
we can redefine the fields by a time dependent R-transformation:
\eqn\redfin{\eqalign{\phi^i&\rightarrow e^{-{i\over r} q_i t}\phi^i
\cr  \psi^i_{\alpha}&\rightarrow e^{-{i\over r} (q_i-1) t}\psi^i_{\alpha} \cr F^i& \rightarrow e^{-{i\over r}(q_i-2) t}F^i ~.} }
Translations along $\R$ are then generated by  
\eqn\newH{H = P_0+ {1\over r} R }
and the superalgebra becomes
\eqn\algspcb{\eqalign{&\{Q_{\alpha},\bar Q_{\dot\alpha}\}=2 \sigma^{0}_{\alpha \dot\alpha}\Big( H-{1\over r} R\Big)+{2\over r}\sigma^i_{\alpha \dot \alpha} J^i_l\cr
&[H,Q_{\alpha}]=0~.}}

Equivalently, instead of performing the redefinition \redfin, we can turn on a ``pure gauge'' background $U(1)_R$ gauge field
\eqn\backA{A_0={1\over r}~.}
This will be useful below.

The Lagrangian is then given by ($q_{\bar i}=-q_i \delta_{i \bar i}$):
\eqn\genkahs{\eqalign{
&\CL_{\S^3\times \R}^{B}=K_{i\bar j}\left(F^i \bar F^{\bar j} +D_t\phi^i D_t{\bar \phi}^{\bar j}-\partial_a \bar \phi^{\bar j}\partial^a \phi^i\right)+ F^i \bar W_i +\bar F^{\bar j} \bar W_{\bar j}\cr
&\qquad \qquad-{i\over r}K_i D_t \phi^i+ {i\over r} K_{\bar j}D_t \bar \phi^{\bar j} \cr
&\CL_{\S^3\times \R}^{F}=-{i} K_{i \bar j}\Big(\bar \psi^{\bar j} \bar \sigma^0 D_t \psi^i+\bar \psi^{\bar j} \bar \sigma^a D_a\psi^i\Big)-{1\over 2} W_{i j} \psi^i \psi ^j -{1\over 2} \bar W_{\bar i \bar j} \bar \psi^{\bar i} \bar \psi^{\bar j}\cr
&\qquad\qquad -{1\over 2}K_{i j \bar j} \bar F^{\bar j} \psi^i \psi^j -{1\over 2}K_{\bar i \bar j  j}  F^{ j} \bar \psi^{\bar i} \bar \psi^{\bar j}+{1\over 4}K_{ i j \bar i \bar j} \psi^i \psi^j \bar \psi^{\bar i}\bar \psi^{\bar j}\cr
&D_t\phi^i=\Big(\partial_t-{i\over r} q_i\Big)\phi^i,\;\;\;\;\;D_t\bar \phi^{\bar i}=\Big(\partial_t - {i\over r} q_{ \bar i}\Big)\bar \phi^{\bar i}\cr
& D_t\psi^i=\left(\partial_t- {i\over r}\Big(q_i-{1\over 2}\Big)\right)\psi^i+\Gamma^{i}_{j l}\psi^j D_t\phi^l\cr
&D_a \psi^i=\grad_a\psi^i+\Gamma^{i}_{j l}\psi^j \partial_a \phi^l.}}
As before, the Lagrangian is invariant (up to a total derivative) under R-invariant K\"ahler transformations $K\rightarrow K+Y+\bar Y$ satisfying $\sum_i q_i Y_i \phi_i=0$.

All the terms of order ${1\over r}$ in \genkahs\ are given by\foot{In our conventions when gauging a conserved current $j^{\mu}$ we add to the Lagrangian $-A_{\mu}j^{\mu}$}
\eqn\leadostr{{1\over e} \CL_{\S^3\times \R}^{(1)} =-{3\over 2}J^{FZ}_0+J^R_0 ~,}
where $J_{\mu}^R$ is the conserved R-current
\eqn\jzeror{J^R_{\mu}= -i K_{i \bar j} q_{ \bar j} \bar \phi^{\bar j} \partial_{\mu} \phi^i-i K_{i \bar j} q_i \phi^i \partial_{\mu} \bar \phi^{\bar j}+K_{i \bar j} (q_i-1)\bar \psi^{\bar j} \bar \sigma_{\mu} \psi^i+K_{ i l \bar j} \bar \psi^{\bar j} \bar \sigma_{\mu} \psi^i q_l \phi^l.}

The parameters of the flat space theory were constrained to be Poincar\'e invariant.  Given that this symmetry is broken to $SU(2)_l \otimes SU(2)_r \otimes U(1)$, there are additional parameters we can turn on.  These can be thought of as background fields.  Of particular interest are background gauge fields associated with the global symmetry of the theory.  For every global non-R-symmetry $U(1)_s$ there is a conserved current $j_\mu^{s}$ and charge $Q^s$.  Then we can add background gauge fields $a^s_\mu $ by coupling them to the currents and adding appropriate seagull terms which are quadratic in $a_\mu^s$.   We turn on background gauge fields which preserve the $SU(2)_l \otimes SU(2)_r \otimes U(1)$ isometry
\eqn\asbackground{a^s_0 = {v_s \over r} ~,}
where $v_s$ are dimensionless real constants.  Denoting by $q_{is}$ the $U(1)_s$ charge of $\phi^i$, this background gauge field has the effect of changing the parameters $q_i$ in \redfin\genkahs\jzeror\ as
\eqn\changeqi{q_i \to q_i + \sum_s q_{is} v_s~.}
Such background fields will play an important role below.

\newsec{New minimal SUGRA}

Starting from \partsugra\ and giving an expectation value to $b_0$ we realized the need for an R-symmetry and for a background $U(1)_R$ gauge field in order to have time independent supercharges on $\S^3 \times \R$. This suggests the use of ``new minimal Supergravity"~\refs{\SohniusTP} to analyze this case.

In the presence of an R-symmetry supergravity can be coupled directly to the R-multiplet, which is distinct from the FZ-multiplet.  It contains the R-current \jzeror\ as its lowest component and it satisfies~\refs{\GatesNR} (for a recent discussion see \KomargodskiRB):
\eqn\Rmultiplet{\bar D^\alphadot \CR_{\alpha \alphadot } = \chi_\alpha \qquad ;\qquad \bar D_\alphadot \chi_\alpha=0\qquad ; \qquad D^\alpha \chi_\alpha=\bar D_\alphadot \bar \chi^\alphadot.}
$\chi^{\alpha}$ satisfies the equations of a chiral field strength and its component expansion
\eqn\defchi{\chi_{\alpha}=-i \lambda_{\alpha}+\Big(\delta^{\beta}_{\alpha}D+2i \sigma^{\mu}\bar \sigma^{\nu}(\partial_{\mu}\CA_{\nu}-\partial_{\nu}\CA_{\mu})\Big)+...}
contains a vector $\CA_{\mu}$.  For a WZ-model
\eqn\chiWZ{\chi_\alpha=\bar D^2 D_\alpha U \qquad ; \qquad U=  K-{3\over 2}\sum q_i\phi^i K_i}
and
\eqn\Athbth{\eqalign{
\CA_{\mu}= U|_{\bar \theta \bar \sigma^{\mu}\theta}&=-i K_i \partial_{\mu} \phi^i+i K_{\bar j} \partial_{\mu} \bar \phi^{\bar j}-{3\over 2}i K_{i \bar j} q_{\bar  j} \bar \phi^{\bar j} \partial_{\mu} \phi^i-{3\over 2}i K_{i \bar j} q_{i} \phi^{ i} \partial_{\mu}\bar  \phi^{\bar j}\cr
&-K_{i \bar j} \bar \psi^{\bar j} \bar \sigma_{\mu} \psi^i+{3\over 2} K_{i \bar j} q_i \bar \psi^{\bar j}\bar \sigma_{\mu}\psi^i+{3\over 2} K_{i l \bar j} q_l \phi^l  \bar \psi^{\bar j}\bar \sigma_{\mu}\psi^i ~.}}
As in the FZ-multiplet, the R-multiplet \Rmultiplet\ is not unique.  It can be improved by shifting the R-current by any conserved global current.  This amounts to changing the values of $q_i$ in \jzeror\chiWZ.

There are two real auxiliary fields in the ``new minimal" gravity multiplet: $A^{\mu}$ and a conserved\ $V^{\mu}={1\over 4}\epsilon^{\mu\nu\rho\lambda}\partial_{\nu}B_{\rho\lambda}$ (terms proportional to the gravitino are set to zero). Taking the $M_p\rightarrow \infty$ limit we get the following Lagrangian for the matter fields~\refs{\SohniusFW}:

\eqn\newmin{\eqalign{
&{1\over e} \CL^{B}=\left({1\over 2}  \CR-3  V_{\mu}V^{\mu}\right)\left({1\over 4} K_i q_i \phi^i -{1\over 4} K_{\bar i} q_{\bar i}\bar \phi^{\bar i} \right) +K_{i\bar j}\left(F^i \bar F^{\bar j} -D_{\mu}\phi^i D^{\mu}{\bar \phi}^{\bar j}\right)\cr
&\qquad \qquad +{iV^{\mu}}\left(K_i D_{\mu} \phi^i- K_{\bar j}D_{\mu} \bar \phi^{\bar j}\right)+ F^i \bar W_i +\bar F^{\bar j} \bar W_{\bar j} \cr
&{1\over e} \CL^{F}=-{i} K_{i \bar j}\bar \psi^{\bar j} \bar \sigma^\mu D_\mu \psi^i-{1\over 2} W_{i j} \psi^i \psi ^j -{1\over 2} \bar W_{\bar i \bar j} \bar \psi^{\bar i} \bar \psi^{\bar j}\cr
&\qquad\qquad -{1\over 2}K_{i j \bar j} \bar F^{\bar j} \psi^i \psi^j -{1\over 2}K_{\bar i \bar j  j}  F^{ j} \bar \psi^{\bar i} \bar \psi^{\bar j}+{1\over 4}K_{ i j \bar i \bar j} \psi^i \psi^j \bar \psi^{\bar i}\bar \psi^{\bar j}\cr
&D_{\mu}\psi^i=\Big(\grad_{\mu}- i(q_i-1)A_{\mu}-{i\over 2}V_{\mu}\Big)\psi^i+\Gamma^{i}_{j l}\psi^j D_\mu\phi^l\cr
&D_\mu \phi^i=(\partial_{\mu} -{iq_i A_{\mu}})\phi^i,\qquad D_\mu \bar \phi^{\bar i}=(\partial_{\mu} -{iq_{ \bar i} A_{\mu}})\bar \phi^{\bar i} }}

The Lagrangian is invariant under local R-symmetry transformations parameterized by $\Lambda(x)$  under which $A_{\mu}\rightarrow A_{\mu}+\partial_{\mu} \Lambda$.  The terms linear in the auxiliary fields are easily recognized as
\eqn\linearan{ V^{\mu}\left({3\over 2} J^R_{\mu} -\CA_{\mu}\right)-A^{\mu} J_{\mu}^R=~ {3\over 2}V^{\mu} J^{FZ}_{\mu}-A^{\mu} J_{\mu}^R~.}
As we have emphasized a number of times above, this expression in terms of the currents is more general than the example of WZ-models we have been discussing.

In a superconformal theory $D^\alphadot \CR_{\alpha \alphadot }=0$ and hence $\CA_{\mu}=0$.  Therefore, the terms proportional to $\CA_{\mu}$ in \linearan\ are a measure of the violation of conformality.

The variations of the chiral superfields components are~\refs{\SohniusFW}
\eqn\varschm{\eqalign{
&\delta \phi^i= -\sqrt{2} \zeta \psi^i\cr
&\delta \psi^i_{\alpha}= -\sqrt{2} \zeta_{\alpha} F^i-i \sqrt{2} (\sigma^{\mu}\bar \zeta)_{\alpha} (\partial_{\mu}-i q_i A_{\mu}) \phi^i \cr
&\delta F^i=-i \sqrt{2} \bar \zeta \bar \sigma^{\mu} \Big(\grad_{\mu} -i (q_i-1)A_{\mu}-{i\over 2} V_{\mu} \Big)\psi^i }}
and the gravitino variation is:
\eqn\gravvar{\eqalign{&\delta \psi_{\mu}^{\alpha}=-2 \grad_{\mu} \zeta^{\alpha}-2 i V^{\nu} ( \zeta\sigma_{\nu\mu})^{\alpha} -2 i(V_{\mu}-A_{\mu})\zeta^{\alpha} ~,\cr
&\delta\bar \psi_{\mu \alphadot}=-2 \grad_{\mu} \bar \zeta_{\alphadot}+2 i V^{\nu} (\bar  \zeta\bar \sigma_{\nu\mu})_{\alphadot} +2 i(V_{\mu}-A_{\mu})\bar \zeta_{\alphadot}}}
As in \susycon\ we view $V_{\mu}$ and $A_{\mu}$ as complex vectors.
The conditions stemming from \gravvar\ requiring four unbroken supercharges are:
\eqn\finalsolnmin{\eqalign{
&\grad_{\mu}V_{\nu} =0\cr
&\partial_{[\mu} A_{\nu]}=0 \cr
&W_{\mu\nu\kappa\lambda}=0\cr
&\CR_{\mu\nu} =-{2} (V_\mu V_\nu - g_{\mu \nu} V_\rho V^\rho)}}

We can find nontrivial $\zeta$ such that \gravvar\ vanishes for $\S^3 \times \R$ by setting
\eqn\newma{V_i=A_i=0 \qquad , \qquad V_0={1\over r}}
where $r$ is the radius of the sphere. The nonzero value of $V^0={1\over 4} \epsilon^{0ijk} \partial_i B_{jk}$ can be interpreted as nonzero flux of $H=dB$ through our $\S^3$. 

The value of $A_0$ is arbitrary and by changing it we obtain Lagrangians related by redefinitions like \redfin.  Three cases have a natural interpretation:
\item{1.} $A_0=V_0$ results in a time independent $\zeta$ and conserved supercharges; it gives \genkahs\ and the corresponding superalgebra  \algspcb.
\item{2.} $A_0=0$ gives the Lagrangian obtained directly in the ``old minimal" formalism  \extrab\ with $\zeta$ satisfying \grazer; the superalgebra is given by \algspc.
\item{3.}
Finally for $A_0={3\over 2}V_0$ the superalgebra is:
\eqn\algspcd{\eqalign{
&\{Q_{\alpha},{Q^{\dagger}}_{\dot\alpha}\} ={2\over r} \sigma^{0}_{\alpha \dot\alpha}\Big( \Delta- {3\over 2 }R\Big)+{2\over r}\sigma^i_{\alpha \dot \alpha} J^i_l\cr
&[\Delta,Q_{\alpha}]=-{1\over 2} Q_{\alpha}.}}
where $\Delta$ generates translations along $\R$.
For a superconformal theory $\Delta$ can be identified with the dilatation generator in the superconformal algebra. The ${1\over r}$ terms in the Lagrangian are given by $\CA_0={3\over2}J_0^R -{3\over 2}J_0^{FZ}$, which indeed vanishes for a SCFT.

\bigskip
Finally, we would like to emphasize another consequence of the use of the new-minimal formalism.  Some rigid supersymmetric theories do not have an FZ-multiplet \KomargodskiRB.  These are theories in which the superfield $U$ in \chiWZ\ is not well defined.  This happens either when the theory has nonzero FI-terms or when the K\"ahler form of $K$ is not exact.  Such theories can be coupled to the old minimal set of auxiliary fields only if they have an R-symmetry.  This is most easily done in the new-minimal formalism.  Indeed, it is straightforward to check that the Lagrangian \newmin\ or the more abstract presentation of the terms of order $1\over r$ \linearan\ are are well defined even when $U$ is not (to do that, integrate by parts the term proportional to $V^\mu = {1\over 4} \epsilon^{\mu \nu\rho \sigma} \partial_\nu B_{\rho \sigma}$).

\newsec{$\S^3\times \S^1$}

In this section we discuss the theory on $\S^3\times \S^1$.
We start by analyzing the analytic continuation of our Lorentzian theory on $\S^3\times \R$ to Euclidean signature.  The analytic continuation of the flat space theory is standard.  But what should we do with the various background fields?  Recall that in the old minimal presentation we used background $b_0$ \bzback\ and a background $U(1)_R$ gauge field $A_0$ \backA\ and in the new minimal formalism we used background $V_0$ and $A_0$ \newma.  Furthermore, we also faced the freedom to turn on background gauge fields for non-R-symmetries $a_0$ \asbackground.

The conditions for unbroken supersymmetry have led us in the Lorentzian theory to backgrounds satisfying
\eqn\CRbV{\CR={2\over 3} b_\mu b^\mu =6 V_\mu V^\mu <0 ~.}
Denoting the Euclidean time direction by $4$, this suggests that we should take \eqn\bVe{b_4 =- 3V_4={3i\over r} ~.}
We recall that we needed such imaginary values of the background auxiliary fields also in the case of $\S^4$ \supcon.  The situation with the background $U(1)_R$ gauge field $A_0$ is similar.  It was needed in order to make the supercharges independent of Lorentzian time.  If we want them to be independent of Euclidean time we should take
\eqn\Aeu{A_4=-{i\over r} }
i.e.\ it should also be imaginary.  Finally, let us discuss the background non-R-gauge fields $a^s$.  Analogy with \changeqi\ suggests that we should take
\eqn\aeu{a^s_4=-{iv_s\over r} }
with real $v_s$.  However, we will see below that it makes sense to consider complex $v_s$ in \aeu.
The Euclidean Lagrangian denoted by $\S^3\times\R_E$ is given by\foot{The Lagrangian is written in terms of the $SU(2)_l$ doublets  $\psi_{E\alpha}=\psi_{\alpha}$ and $\bar \psi_{E\alpha}=i \sigma^4_{\alpha \alphadot}\bar \psi^{\alphadot} $ and we suppress the subscript $E$.  These are contracted with $\epsilon_{\alpha \beta}$ so that, e.g. $\bar \psi \sigma^i \psi=\bar \psi_{\alpha} \epsilon_{\alpha \beta} \sigma^i_{\beta\gamma} \psi_{\gamma}$ and $\overline{\psi \psi}=-\bar \psi\bar\psi$. We also used $\bar \sigma_4=\sigma_4=- i \unit$ and $\bar\sigma^a=-\sigma^a$.}:
\eqn\StimesRE{\eqalign{
&\CL_{\S^3\times \R_E}^{B}=K_{i\bar j}\left(D_{4}\phi^i D_{4}{\bar \phi}^{\bar j}+\partial_a \bar \phi^{\bar j}\partial^a \phi^i-F^i \bar F^{\bar j}\right)- F^i \bar W_i -\bar F^{\bar j} \bar W_{\bar j}\cr
&\qquad\qquad \qquad-{1\over r}K_i D_4 \phi^i+ {1\over r} K_{\bar j}D_4 \bar \phi^{\bar j} \cr
&\CL_{\S^3\times \R_E}^{F}=-K_{i \bar j}\Big( \bar \psi^{\bar j}  D_4 \psi^i-i\bar \psi^{\bar j}  \sigma^a D_a\psi^i\Big)+{1\over 2} W_{i j} \psi^i \psi ^j -{1\over 2} \bar W_{\bar i \bar j} \bar \psi^{\bar i} \bar \psi^{\bar j}\cr
&\qquad\qquad\qquad +{1\over 2}K_{i j \bar j} \bar F^{\bar j} \psi^i \psi^j -{1\over 2}K_{\bar i \bar j  j}  F^{ j} \bar \psi^{\bar i} \bar \psi^{\bar j}+{1\over 4}K_{ i j \bar i \bar j} \psi^i \psi^j \bar \psi^{\bar i}\bar \psi^{\bar j}\cr
&D_{4}\phi^i=\Big(\partial_{4}-{1\over r} q_i\Big)\phi^i,\;\;\;\;\;D_{4}\bar \phi^{\bar i}=\Big(\partial_{4} - {1\over r} q_{ \bar i}\Big)\bar \phi^{\bar i}\cr
& D_{4}\psi^i=\left(\partial_4- {1\over r}\Big(q_i-{1\over 2}\Big)\right)\psi^i+\Gamma^{i}_{j l}\psi^j D_4\phi^l\cr
&D_a \psi^i=\grad_a\psi^i+\Gamma^{i}_{j l}\psi^j \partial_a \phi^l.}}

Now we are ready to compactify the Euclidean time direction to $\S^1$.  The partition function of this system can be interpreted as a trace over the Hilbert space
\eqn\traH{Z=\Tr (-1)^F \exp\left(-\beta H - {\beta \over r} \sum_s v_s Q_s\right)~.}
Here we used the Hamiltonian $H$ of \algspcb\ which commutes with the supercharges and $Q_s$ is the charge of $U(1)_s$.

If the underlying theory is conformal, \traH\ is known as the conformal index with chemical potentials $v_s$ \MinwallaMA.  But following \refs{\RomelsbergerEG,\RomelsbergerEC,\DolanQI} we can study it also in non-conformal theories.  In that case the term ``superconformal index'' is clearly inappropriate.

The Hilbert space is in representations of
\eqn\suysgropu{SU(2|1)_l\otimes SU(2)_r \otimes U(1)_R \otimes_s U(1)_s}
and the objects in the exponent of \traH\ commute with all the elements of this supergroup.  The long representations of this group do not contribute to this trace.  The short representations are constructed out of a highest weight state with $P_0 = {2\over r}j$ where $P_0$ is the generator of $U(1) \subset SU(2|1)_l$ (see \algspc) and $j$ is the quantum number of $SU(2)_l \subset SU(2|1)_l$.  Such short representations contribute to the trace \traH\ $\pm \exp\left[-{\beta\over r}\left(2 j + R + \sum_s v_s Q^s\right)\right]$, where $R$ is the R-charge of the highest weight state and $Q^s$ are the $U(1)_s$ charges of the states in the representation \refs{\RomelsbergerEG,\RomelsbergerEC}.

We note that we could have also added to the trace \traH\ additional chemical potentials without ruining its nice properties.  Some of them do not respect the isometry of the sphere and correspond to squashing it.  We will not do it here.

It is important that the values of $r P_0=2 j$ of the short representations are quantized.  Therefore, the values of $H=P_0 + {1\over r}R$ and of $Q_s$ of the states in these representations cannot depend on the parameters of the theory and on renormalization group flow.  Hence, the $\S^3\times \R$ partition function $Z$ \traH\ is independent of the parameters of the Lagrangian and the renormalization group scale and depends only on the dimensionless parameters $\beta \over r$ and $v_s$.  Equivalently, the parameters in the flat space Lagrangian multiply operators, which are given by commutators with the supercharge.  Therefore, their expectation values must vanish and $Z$ does not depend on them.
This fact has allowed \refs{\RomelsbergerEG,\RomelsbergerEC,\DolanQI-\DolanRP} to compute $Z$ in many interesting cases.

We will find it useful to extend the previous discussion to complex $v_s$.  The real parameters ${\rm Re}\,  v_s$ have the effect of shifting the R-charges \changeqi.  The imaginary parts ${\rm Im}\,  v_s$ also have a natural interpretation.  If we view the theory on $\S^3$ as a three dimensional field theory, then $m_s = {1\over r} {\rm Im}\,  v_s$ can be interpreted as ``real mass terms.''  In the next section we will discuss theories on $\S^3$ in more detail.  Here we will simply comment that with such complex $v_s$ the dependence of $Z$ on $v_s$ is holomorphic.

\newsec{$\S^3$}

We now turn to consider three-dimensional theories on $\S^3$.  For simplicity we will focus on theories obtained by taking a four-dimensional theory on $\S^3\times \S^1$ in the limit that the circumference of the circle goes to zero $\beta \to 0$ (See the recent papers~\refs{\GaddeIA,\ImamuraUW} for  related discussions); but as will be clear, our conclusions are not limited to such theories.

Starting with \traH\ and taking $\beta \to 0$ with an appropriate limit of the Lagrangian parameters we find a three-dimensional $\CN=2$ theory with a global $U(1)_R$ symmetry on $\S^3$ with Lagrangian (recall that $q_{\bar i}=-\delta_{i \bar i} q_i$)
\eqn\SEuc{\eqalign{
&\CL_{\S^3}^{B}=K_{i\bar j}\left(\partial_a \bar \phi^{\bar j}\partial^a \phi^i+{1\over r^2}q_i q_{\bar j} \phi^{i}\bar \phi^{\bar j}-F^i \bar F^{\bar j}\right)- F^i \bar W_i -\bar F^{\bar j} \bar W_{\bar j}\cr
&\qquad \qquad+{1\over r^2}q_i K_i \phi^i- {1\over r^2}q_{\bar j} K_{\bar j} \bar \phi^{\bar j} \cr
&\CL_{\S^3}^{F}={i}K_{i \bar j}\Big(\bar \psi^{\bar j} \sigma^a D_a\psi^i-{i\over r}\Big(q_i-{1\over 2}\Big) \bar \psi^{\bar j} \psi^i -{i\over r} q_l\Gamma^{i}_{j l} \bar \psi^{\bar j} \psi^j \phi^l\Big)+{1\over 2} W_{i j} \psi^i \psi ^j -{1\over 2} \bar W_{\bar i \bar j} \bar \psi^{\bar i} \bar \psi^{\bar j}\cr
&\qquad\qquad +{1\over 2}K_{i j \bar j} \bar F^{\bar j} \psi^i \psi^j -{1\over 2}K_{\bar i \bar j  j}  F^{ j} \bar \psi^{\bar i} \bar \psi^{\bar j}+{1\over 4}K_{ i j \bar i \bar j} \psi^i \psi^j \bar \psi^{\bar i}\bar \psi^{\bar j}\cr
&D_a \psi^i=\grad_a\psi^i+\Gamma^{i}_{j l}\psi^j \partial_a \phi^l.}}
The Lagrangian of this theory depends on the parameters of the flat space Lagrangian as well as on the complex dimensionless parameters $v_s$ introduced via \changeqi.  As we remarked above, $m_s={1\over r} {\rm Im}\,  v_s$ can be interpreted as ``real mass terms'' in the three-dimensional theory, while ${\rm Re}\, v_s$ are parameters that determine how the theory is placed on $\S^3$, through shifts of the R-current.

For generic couplings the theory on $\S^3$ is not reflection positive and does not correspond to a unitary Lorentzian theory on $dS_3$.  This fact is similar to our discussion above about $\S^4$ and is easily visible in the terms of order $1\over r$ in the Lagrangian.  Using our construction, which is based on background fields, this lack of reflection positivity arises because of the complex values of in \bVe\Aeu\aeu.

The discussion above easily leads to the following conclusions about this three dimensional theory.

First, as in the discussion about $\S^3\times \S^1$, the $\S^3$ partition function $Z$ is independent of most of the parameters in the flat space Lagrangian on $\R^3$.  It depends only on the real mass terms $m_s = {1\over r} {\rm Im}\,  v_s$.  This fact has made the computations in \refs{\KapustinXQ-\MinwallaMA} possible.

Second, the dependence on $v_s$ is holomorphic.  This fact might seem strange and was referred to as ``mysterious'' in \JafferisUN, because the mass terms $m_s = {1\over r} {\rm Im}\,  v_s$ are parameters in the flat space $\R^3$ Lagrangian, while ${\rm Re}\,  v_s$ affect the choice of the R-current.  Constructing these theories by coupling them to background fields makes this holomorphy manifest.  In particular, the complex number $v_s$ is a background field that couples to the scalar operator $j_s$ in the $\bar\theta \theta$ component of the supersymmetry multiplet of the conserved $U(1)_s$ current.

\bigskip
\centerline{\bf Acknowledgements}
We would like to thank D.~Gaiotto, J.~Gauntlet, C.~Hull,  G.~Moore, J.~Maldacena, M.~Rocek, and E.~Witten for useful discussions. We are particularly thankful to D.~Jafferis, Z.~Komargodski, and D.~Shih for participation in the early stages of this project.  NS thanks the Simons Center for Geometry and Physics for its kind hospitality.
The work of GF was supported in part by NSF grant PHY-0969448. The work of NS was supported in part by DOE grant
DE-FG02-90ER40542.

\listrefs

\end